\PassOptionsToPackage{svgnames, x11names, table}{xcolor} 

\documentclass{article} 

\usepackage[preprint]{colm2026_conference}

\usepackage{microtype}
\usepackage{hyperref}
\usepackage{url}
\usepackage{booktabs}
\usepackage{lineno}
\usepackage{graphicx}
\usepackage{float}       
\usepackage{amsmath}
\usepackage{multirow}
\usepackage{tabularx}
\usepackage{makecell}
\usepackage{fontawesome5}
\usepackage{pifont}
\usepackage{listings}
\usepackage{inconsolata} 

\usepackage{tcolorbox}
\tcbuselibrary{listings, skins, breakable}

\usepackage{wrapfig}

\newtcolorbox{boxL}{
    fontupper = \color{black},
    rounded corners,
    arc = 6pt,
    colframe = black!50, 
    boxrule = 0pt, 
    bottomrule = 4.5pt ,
    breakable,
}

\definecolor{yescolor}{RGB}{0, 150, 0}
\definecolor{nocolor}{RGB}{200, 200, 200}
\newcommand{\yes}{\textcolor{yescolor}{\ding{51}}}
\newcommand{\no}{\textcolor{nocolor}{\ding{55}}}
\definecolor{summarybg}{gray}{0.95}
\definecolor{oursbg}{RGB}{235, 245, 255}

\definecolor{darkblue}{rgb}{0, 0, 0.5}
\hypersetup{colorlinks=true, citecolor=darkblue, linkcolor=darkblue, urlcolor=darkblue}

\title{\texttt{AppellateGen:} A Benchmark for Appellate Legal Judgment Generation}

\author{
\textbf{Hongkun Yang\textsuperscript{1,2}\thanks{Equal contribution. Work done while Hongkun Yang was an intern at Nanyang Technological University.}},
\textbf{Lionel Z. Wang\textsuperscript{1,3}\thanks{Equal contribution. Corresponding author}},
\textbf{Wei Fan\textsuperscript{4}},
\textbf{Yiran Hu\textsuperscript{5}},
\textbf{Lixu Wang\textsuperscript{1}},\\
\textbf{Chenyu Liu\textsuperscript{3}},
\textbf{Yu Zeng\textsuperscript{6}},
\textbf{Shenghong Fu\textsuperscript{3}},
\textbf{Lei Gong\textsuperscript{3}},
\textbf{Zhengxin Zhang\textsuperscript{7}},\\
\textbf{Haoyang Li\textsuperscript{3}},
\textbf{Jiexin Zheng\textsuperscript{6}},
\textbf{Xin Xu\textsuperscript{3}}\\[1ex]
\textsuperscript{1}Nanyang Technological University,
\textsuperscript{2}Ocean University of China,\\
\textsuperscript{3}The Hong Kong Polytechnic University,
\textsuperscript{4}Hong Kong University of Science and Technology,\\
\textsuperscript{5}Tsinghua University,
\textsuperscript{6}University of Science and Technology of China,\\
\textsuperscript{7}Cornell University
}

\begin{document}

\ifcolmsubmission
\linenumbers
\fi

\maketitle

\begin{abstract}
Legal judgment generation is a critical task in legal intelligence. However, existing research in legal judgment generation has predominantly focused on first-instance trials, relying on static fact-to-verdict mappings while neglecting the dialectical nature of appellate (second-instance) review. To address this, we introduce \texttt{AppellateGen}, a benchmark for second-instance legal judgment generation comprising 7,351 case pairs. The task requires models to draft legally binding judgments by reasoning over the initial verdict and evidentiary updates, thereby modeling the causal dependency between trial stages. We further propose a judicial Standard Operating Procedure (SOP)-based Legal Multi-Agent System (SLMAS) to simulate judicial workflows, which decomposes the generation process into discrete stages of issue identification, retrieval, and drafting. Experimental results indicate that while SLMAS improves logical consistency, the complexity of appellate reasoning remains a substantial challenge for current LLMs. The dataset and code are publicly available at: \url{https://anonymous.4open.science/r/AppellateGen-5763}.
\end{abstract}

\section{Introduction}

The integration of Large Language Models (LLMs) into the legal domain has advanced legal NLP, moving from simple clause retrieval to complex reasoning tasks such as passing the Bar Exam \citep{katz2024gpt}, and transitioning from discriminative Legal Judgment Prediction (LJP) \citep{cui2023chatlaw, feng2022legal} to Legal Document Generation \citep{li2025casegen, yue2023disc}. Despite these advancements, contemporary methodologies predominantly adopt a First-Instance Perspective \citep{zhong2018legal,liu2023ml, xu2020distinguish}, where generative models synthesize judgments directly from static factual descriptions. However, this linear approach neglects a critical phase: the Appellate (Second-Instance) Review (illustrated in \textbf{Appendix~\ref{app: appellate}}). Distinct from the fact-to-verdict mapping of a first trial, appellate proceedings are inherently dialectical, necessitating a comparative analysis of the original judgment against appellant grievances to rectify errors in fact-finding or legal application \citep{wechsler1977appellate}.

This distinction is pronounced in Civil Law jurisdictions. Here, appellate courts operate under the principle of \textit{continuation of trial}, allowing for the re-examination of facts and the admission of new evidence \citep{apple1995primer}. This procedural feature necessitates models capable of processing dynamic factual updates. According to judicial statistics from the Supreme People's Court of China, the reversal rate of second-instance cases consistently hovered around 10\% to 11\% (exceeding 637,000 cases) from 2021 to 2023 \citep{SPC2021, SPC2022, SPC2023}. This substantial volume of modified verdicts underscores the complexity inherent in appellate adjudication \citep{luneburg1981specially,bruhl2010deciding}, which demands complex reasoning to identify discrepancies and produce a binding judgment that affirms, reverses, or remands the initial ruling.

Despite the importance of appellate review, existing legal benchmarks exhibit limitations in \textbf{Task Formulation} and \textbf{Contextual Granularity}. First, regarding formulation, existing benchmarks primarily assess discriminative capabilities through multiple-choice questions or classification \citep{chalkidis-etal-2022-lexglue, fei-etal-2024-lawbench}. While recent initiatives have integrated generative tasks \citep{li-etal-2025-legalagentbench, li2024lexeval}, they predominantly focus on drafting documents from static facts, overlooking the dialectical nature of appeals. Second, in terms of granularity, current methodologies typically model legal documents as isolated data points. Retrieval-centric benchmarks \citep{pipitone2024legalbench} and document generation datasets \citep{li2025casegen} generally treat case processing as a single-step inference task. This approach neglects the procedural interdependency between adjudicatory stages, specifically the dependency between initial judgments and appellate reviews.

To bridge this gap, we introduce \texttt{AppellateGen}, the first benchmark dedicated to \textbf{Second-Instance Legal Judgment Generation}. Unlike previous benchmarks, \texttt{AppellateGen} models the complete dispute lifecycle. We curate \textbf{7,351 paired cases}, linking first-instance judgments with their corresponding appellate outcomes. We enhance this with a fine-grained annotation schema that identifies points of contention, extracts new facts introduced during the appeal, and categorizes the rationale for reversal. To validate the challenge posed by \texttt{AppellateGen} and investigate robust solutions, we propose the \textbf{judicial Standard Operating Procedure (\underline{S}OP)-based \underline{L}egal \underline{M}ulti-\underline{A}gent \underline{S}ystem (SLMAS)} as a strong baseline. We observe that standard LLMs often struggle when processing the conflicting narratives and long-context evidence inherent in appellate review, leading to hallucinations. To address this, SLMAS imposes a Standard Operating Procedure (SOP) inspired by judicial workflows, decomposing the generation task into discrete stages: \textit{Analysis, Search, Predict, and Write}. By orchestrating agents specialized in issue identification, retrieval, prediction, and drafting, our framework explicitly models the intermediate reasoning states required for valid adjudication. Empirical results demonstrate that this modular approach significantly enhances logical consistency and reversal prediction accuracy, surpassing both commercial and domain-specific baselines.

In summary, our contributions are threefold:

\begin{itemize}
    \item \textbf{Pioneering Task Definition:} We extend the frontier of legal NLP from static first-instance judgment prediction to Second-Instance Legal Judgment Generation. This task necessitates a paradigm shift towards \textit{dynamic dialectical reasoning}, requiring generative models to identify judicial errors and synthesize conflicting narratives between the original verdict and new evidence.
    
    \item \textbf{The \texttt{AppellateGen} Benchmark:} We construct the first large-scale benchmark of 7,351 paired legal cases that explicitly links trial proceedings with appellate outcomes. Equipped with a fine-grained annotation schema for reversal rationales and evidentiary updates, this dataset serves as a rigorous testbed for evaluating long-context reasoning and causal consistency in legal adjudication.
    
    \item \textbf{Methodological Framework \& Insights:} We propose a judicial Standard Operating Procedure (\underline{S}OP)-based \underline{L}egal \underline{M}ulti-\underline{A}gent \underline{S}ystem (SLMAS). By decomposing the generation process into dispute identification, retrieval, prediction, and drafting, SLMAS has more outstanding generation performance and mitigates logical hallucinations. Furthermore, our extensive evaluation reveals that general-purpose LLMs with strong reasoning capabilities outperform traditional legal domain-specific models on this complex task.
\end{itemize}

\section{Related Work}

\subsection{From Legal Prediction to Appellate Generation via Multi-Agents}

Legal Judgment Prediction (LJP) has progressed significantly, evolving from hierarchical structures \citep{luo2017learning} and few-shot learning \citep{hu2018few} to deploying attention-based architectures \citep{chen2019charge, li2019mann, kang2019creating} and LLMs \citep{nigam2024rethinking}. Recently, the field has witnessed two major paradigm shifts to address complex legal tasks. First, there is a pivot from mere outcome prediction to full document generation, exemplified by initiatives like CaseGen \citep{li2025casegen}. Second, to address single-agent limitations such as hallucinations, Multi-Agent Systems (MAS) have been introduced through collaborative simulation \citep{ji2023towards, li2023camel}. Frameworks like ChatLaw \citep{cui2023chatlaw} and logic-graph-enhanced models \citep{yuan2026multi} now leverage such role-playing collaboration to optimize judgment prediction and legal consultation. Despite these advancements, existing generative models and multi-agent frameworks share a critical limitation: they are inherently designed for first-instance trials. To bridge this gap, we introduce the task of Second-instance Legal Judgment Generation.

\begin{figure*}[htbp!]
\centering
\begin{tcolorbox}[
    enhanced,
    arc=2pt,
    boxrule=0.8pt,
    colback=white,
    colframe=gray!60,
    colbacktitle=gray!15,
    coltitle=black,
    attach boxed title to top center={yshift=-10pt},
    boxed title style={colframe=gray!60, colback=gray!15}
]
\tiny
\begin{tabularx}{\textwidth}{X|X}
\rowcolor{oursbg}
\textbf{\faGavel \quad First Instance} & \textbf{\faBalanceScale \quad Second Instance} \\ \midrule

\textbf{Full Text:} \par 
\begin{tabular}{p{0.45\textwidth}}
\textit{A case concerning a dispute over the right of subrogation between the plaintiff, PICC, and the defendants, Mr. Wen...}
\end{tabular} & 
\textbf{Full Text:} \par 
\begin{tabular}{p{0.45\textwidth}}
\textit{Mr. Wen appealed the first-instance judgment in a case concerning a subrogation claim dispute...}
\end{tabular} \\ \midrule

\textbf{Facts:} \par 
Wen (fully liable) collided with Han. Han's vehicle was repaired for \$8,586. Insurer paid Han and sought recovery from Wen. Wen claimed "old damages" existed but provided no proof. & 
\textbf{New Facts:} \par 
Court retrieved police body-cam footage. Footage confirms Han admitted to "old injuries" on site. Damage height didn't match the accident impact. \\ \midrule

\textbf{Legal Articles:} \par 
- \textit{Insurance Law} Art. 60 \par 
- \textit{Road Traffic Safety Law} Art. 76 & 
\textbf{Legal Articles:} \par 
- \textit{Civil Procedure Law} Art. 170(1)(2) \\ \midrule

\textbf{Judgment:} \par 
1. PICC (Insurer 2) to pay \$2,000. \par 
2. \textbf{Wen (Defendant) to pay \$6,586.} & 
\textbf{Judgment:} \par 
1. Affirm PICC's \$2,000 payment. \par 
2. \textbf{Revoke Wen's \$6,586 payment order.} \\ \midrule

\rowcolor{summarybg}
\multicolumn{2}{p{0.96\textwidth}}{
    \textbf{Is\_reversal:} \texttt{True} \quad | \quad \textbf{Reason for Reversal:} 
    Based on newly obtained police footage, the repair scope clearly exceeded the damage caused by the current accident. The trial court's \textbf{erroneous fact-finding} regarding the damage amount is corrected, and the excess claim is dismissed.
} \\ \bottomrule

\end{tabularx}
\end{tcolorbox}
\caption{An example from our dataset (translated from Chinese). \textbf{Facts} refers to the key factual findings extracted from the unstructured full text via a Large Language Model (LLM). \textbf{Reason for Reversal} denotes the explanatory rationale synthesized by the LLM, which identifies the critical discrepancies between the first and second instances that led to the judgment change.}
\label{fig:case_id_0_display}
\end{figure*}

\subsection{Benchmarks in the Legal Domain}
Existing benchmarks focus on isolated, single-stage tasks, varying significantly in scope. 
On an international scale, foundational resources like LegalBench \citep{guha2023legalbench} and LexGLUE \citep{chalkidis-etal-2022-lexglue} provide extensive classification annotations, while LEXTREME \citep{niklaus2023lextreme}, FairLex \citep{chalkidis-etal-2022-fairlex}, and ArabLegalEval \citep{hijazi-etal-2024-arablegaleval} extend evaluation to multilingual and fairness contexts. 
In the domain of Chinese law, the CAIL competition series serves as a backbone for recent benchmarks emphasizing cognitive reasoning and syllogism, such as LawBench \citep{fei-etal-2024-lawbench}, LexEval \citep{li2024lexeval}, and LAiW \citep{dai-etal-2025-laiw}. 
Beyond general understanding, recent works have targeted specific system components, including retrieval (LegalBench-RAG \citep{pipitone2024legalbench}, LegalSearchLM \citep{kim2025legalsearchlm}), instruction tuning (LawInstruct \citep{niklaus-etal-2025-lawinstruct}), and agentic workflows (LegalAgentBench \citep{li-etal-2025-legalagentbench}). 
While generative capabilities are partially addressed by CaseGen \citep{li2025casegen} and GreekBarBench \citep{chlapanis2025greekbarbench}, these tasks often lack the dialectical depth of appellate review. A notable exception is the recent AppealCase dataset \citep{huang2025appealcase}, which explicitly targets second-instance scenarios. 
However, this work focuses on the discriminative classification task, failing to capture the complex reasoning required to draft appellate opinions. 

\begin{figure*}
    \centering
    \includegraphics[width=0.8\linewidth]{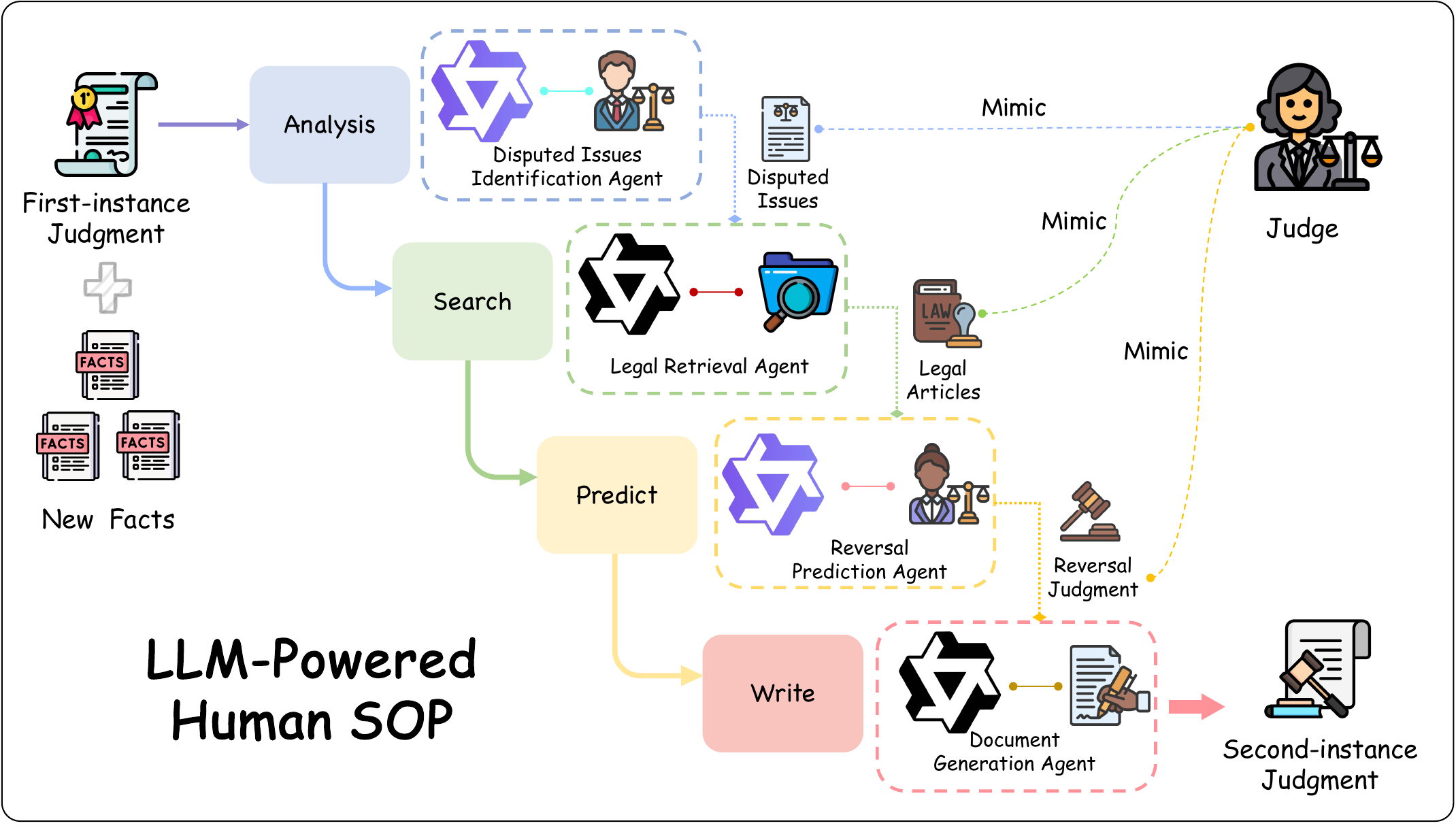}
    \caption{SOP-based Legal Multi-Agent System
(SLMAS). With first-instance judgment and new facts provided, the framework mimics the human judicial Standard Operating Procedure (SOP) by orchestrating four specialized agents. These agents sequentially identify disputed issues, retrieve relevant legal statutes, predict the reversal outcome, and finally draft the appellate judgment, ensuring the generation is grounded in explicit legal reasoning.}
    \label{fig:MAS}
\end{figure*}

\section{\texttt{AppellateGen} Construction}

\subsection{Data Collection and Legal Instances Matching}

We collected 4,706,987 raw legal judgment documents spanning from 2021 to 2024\footnote{This specific timeframe was selected to ensure legal consistency, as the Civil Code and the 11th Amendment to the Criminal Law of the People's Republic of China were enacted in 2021, providing a stable statutory basis for the judgments in our dataset.} from \textit{China Judgments Online}\footnote{\url{https://wenshu.court.gov.cn/}}, the official government repository for Chinese judicial documents. Since the raw repository stores judgments as independent documents without explicit links between first-instance and second-instance proceedings, we developed a multi-stage pipeline to match the same cases in different instances (Details illustrated in \textbf{Appendix \ref{app: Legal Instances Matching}}). To assess the quality of the matching process, we randomly sampled 5\% of the paired cases for manual expert review, achieving a consistency rate of approximately 96\%. After the whole process, all the documents have been de-identified to protect the privacy of litigants. We show an example of our dataset in \textbf{Figure~\ref{fig:case_id_0_display}}.

\subsection{Dataset Statistics}

A final dataset comprising 7,351 high-quality case pairs is curated through the pipeline described above, this sparsity is attributed to three main factors: 1) Natural Filtering: A significant portion of cases conclude at the first instance without appeal; 2) Procedural Outcomes: In the appellate stage, a high volume of cases are resolved through mediation or withdrawal, which typically yield brief procedural rulings rather than the substantive judgments required for reasoning tasks; 3) Strict Quality Control: To ensure the reliability of the benchmark, we enforced rigorous matching criteria based on case numbers and litigant names, discarding pairs with incomplete chains of custody or insufficient textual content. We report the key descriptive statistics in \textbf{Table~\ref{tab:dataset_stats}}. 

\begin{wraptable}{R}{0.5\textwidth}
    \centering
    \setlength{\tabcolsep}{4pt}
    \footnotesize 
    \begin{tabular}{lcc}
    \toprule
    \textbf{Metric} & \textbf{1st Instance} & \textbf{2nd Instance} \\
    \midrule
    Count           & 7,351    & 7,351 \\
    Avg. Length     & 5,004.1  & 5,752.8 \\
    Max. Length     & 40,219   & 41,732 \\
    Min. Length     & 916      & 879 \\
    \midrule
    \textbf{Label Dist.} & \multicolumn{2}{c}{\textit{Rev.: 10.7\% \quad Aff.: 89.3\%}} \\
    \bottomrule
    \end{tabular}
    \caption{Descriptive statistics of the dataset. Lengths are measured in Chinese characters. The dataset contains 7,351 pairs in total (10.7\% Reversal, 89.3\% Affirmation).}
    \label{tab:dataset_stats}
\end{wraptable}

\subsection{Fine-grained Data Annotation}
While standard metrics (e.g., ROUGE \citep{lin2004rouge}) assess lexical overlap, they fail to capture the legal validity and logical consistency required for the task of second-instance document generation. To address this limitation, we designed a fine-grained annotation schema for the matched case pairs. These annotations establish a rigorous semantic ground truth that enables multi-dimensional evaluation beyond surface-level text generation.

Unlike previous datasets that focus solely on static judgment outcomes, our annotation schema captures the \textit{dynamic interaction} and \textit{adjudicative evolution} between the two instances. This allows us to assess whether the generated documents accurately reflect the legal reasoning shifts from the first to the second trial. The annotations were obtained via a hybrid framework combining LLM-driven automated extraction with strict human verification. The schema comprises four key dimensions: \paragraph{Judgment Result Classification:}
This binary label serves as the high-level outcome indicator. We classify the second-instance ruling as either \textit{Affirm} (maintaining the original verdict) or \textit{Reverse} (encompassing cases remanded for retrial or where the judgment was directly altered). This provides the fundamental stance for the generated text.

\paragraph{Rationale for Reversal:}
For cases involving a reversal, we identify the specific legal trigger. This label distinguishes whether the appellate decision was driven by \textit{erroneous fact-finding}, \textit{misapplication of law}, or \textit{procedural violations} in the first instance. This annotation is critical for evaluating if the generated document correctly explains the \textit{why} behind the verdict.

\paragraph{New Evidence and Facts Identification:}
This label summarizes newly ascertained facts derived exclusively from the second-instance judgment, explicitly excluding facts established in the first instance. This design requires the LLM to synthesize the first-instance factual summary with these novel appellate details to perform legal reasoning, thereby simulating authentic judicial practice.

\paragraph{Statutory Citation Extraction:}
This label targets the specific legal articles cited by the appellate court as the controlling authority for its decision. We extract the full citation (e.g., \textit{Article X of the Civil Code}), distinguishing operative laws that dictate the ruling from those merely referenced in the procedural history.


To ensure annotation reliability, we established a robust three-stage pipeline comprising structural document decomposition, LLM-driven annotation with Chain-of-Thought (CoT) prompting, and human-in-the-loop verification. Detailed methodology and evaluation metrics are provided in Appendix \ref{app: Automated Annotation Pipeline}

\section{SOP-based Legal Multi-Agent System}

\subsection{Problem Formulation}
We formalize the task of Second-Instance Legal Judgment Generation as a conditional text generation problem involving multi-hop reasoning. Let $\mathcal{D} = \{(\mathbf{x}_i, \mathbf{y}_i)\}_{i=1}^N$ denote the dataset. For each case, the input context $\mathbf{x}$ comprises a tuple $\mathbf{x} = \langle \mathbf{d}_{\mathrm{1st}}, \mathbf{e}_{\mathrm{new}} \rangle$, where $\mathbf{d}_{\mathrm{1st}}$ represents the textual content of the first-instance judgment (including initial fact-finding and verdict), and $\mathbf{e}_{\mathrm{new}}$ denotes the new evidentiary facts introduced during the appeal (where $\mathbf{e}_{\mathrm{new}} = \emptyset$ if no new evidence is presented).

The target output is the second-instance judgment document, denoted as $\mathbf{y} = (y_1, y_2, \dots, y_T)$. Our objective is to learn a parameterized model $\theta$ that maximizes the log-likelihood of generating $\mathbf{y}$ conditioned on $\mathbf{x}$. Unlike standard end-to-end generation, we posit that the generation of $\mathbf{y}$ strictly depends on a sequence of intermediate reasoning states $\mathbf{z} = \{\mathbf{z}_{\mathrm{issue}}, \mathbf{z}_{\mathrm{law}}, \mathbf{z}_{\mathrm{verdict}}\}$. Thus, the objective function is formulated as:
\begin{equation}
\mathcal{L}(\theta) = \sum_{(\mathbf{x}, \mathbf{y}) \in \mathcal{D}} \log P_\theta(\mathbf{y} \mid \mathbf{x}, \mathbf{z})
\end{equation}
where $\mathbf{z}$ is the latent rationales derived from the judicial Standard Operating Procedure (SOP).

\subsection{Framework Overview}
Directly mapping $\mathbf{x} \to \mathbf{y}$ often leads to hallucinations and logical inconsistencies due to the complexity of appellate review. To address this, we propose a judicial Standard Operating Procedure (\underline{S}OP)-based \underline{L}egal \underline{M}ulti-\underline{A}gent \underline{S}ystem (SLMAS) (Shown in \textbf{Figure \ref{fig:MAS}}) that mimics the human judicial Standard Operating Procedure (SOP) \citep{demarco2013peopleware, hong2023metagpt}.

We model the appellate process as a deterministic assembly line executed by four specialized agents: 
\begin{equation}
\mathcal{A} = \{\mathcal{A}_{\mathrm{issue}}, \mathcal{A}_{\mathrm{retr}}, \mathcal{A}_{\mathrm{pred}}, \mathcal{A}_{\mathrm{write}}\}.
\end{equation}
Each agent $\mathcal{A}_k$ acts as a functional operator that transforms the current state and prior contexts into a structured intermediate output. The overall workflow forms a directed acyclic graph (DAG) of reasoning \citep{digitale2022tutorial}, ensuring that the final generation is grounded in explicit legal logic rather than statistical correlations.

\subsection{Sequential Reasoning Workflow}
The execution of our framework follows a sequential chain corresponding to the four stages of human adjudication: \textit{Analysis, Search, Predict, and Write}.

\paragraph{Stage 1: Disputed Issues Identification (Analysis).}
The process begins with the \textit{Disputed Issues Identification Agent} ($\mathcal{A}_{\mathrm{issue}}$). This agent analyzes the discrepancy between the first-instance judgment and the new evidence to distill the core points of contention. Formally, given the input $\mathbf{x}$, the agent extracts a set of disputed issues $\mathbf{z}_{\mathrm{issue}}$:
\begin{equation}
\mathbf{z}_{\mathrm{issue}} = \mathcal{A}_{\mathrm{issue}}(\mathbf{d}_{\mathrm{1st}}, \mathbf{e}_{\mathrm{new}})
\end{equation}
where $\mathbf{z}_{\mathrm{issue}}$ serves as a structured summary of the appeal grounds, guiding the subsequent retrieval process.

\paragraph{Stage 2: Legal Retrieval (Search).}
To ground the reasoning in statutory law, the \textit{Legal Retrieval Agent} ($\mathcal{A}_{\mathrm{retr}}$) utilizes the identified issues $\mathbf{z}_{\mathrm{issue}}$ to query an external legal knowledge base $\mathcal{K}$ (Details Illustrated in \textbf{Appendix \ref{app: SLMAS}}). Unlike generic retrieval, this agent focuses on statutes specifically relevant to the controversy:
\begin{equation}
\mathbf{z}_{\mathrm{law}} = \mathcal{A}_{\mathrm{retr}}(\mathbf{z}_{\mathrm{issue}}, \mathcal{K})
\end{equation}
Here, $\mathbf{z}_{\mathrm{law}}$ represents the retrieved relevant legal articles and judicial interpretations, providing the normative basis for the review.

\paragraph{Stage 3: Reversal Prediction (Predict).}
Before drafting the document, the \textit{Reversal Prediction Agent} ($\mathcal{A}_{\mathrm{pred}}$) mimics a judge's internal deliberation. This agent integrates the case context and derived legal grounds to adjudicate the appeal, specifically predicting whether the primary verdict warrants affirmation or reversal. The agent generates a predictive rationale $\mathbf{z}_{\mathrm{verdict}}$:
\begin{equation}
\mathbf{z}_{\mathrm{verdict}} = \mathcal{A}_{\mathrm{pred}}(\mathbf{d}_{\mathrm{1st}}, \mathbf{e}_{\mathrm{new}}, \mathbf{z}_{\mathrm{issue}}, \mathbf{z}_{\mathrm{law}})
\end{equation}
This step acts as a logical checkpoint, ensuring the final document's conclusion is consistent with the analyzed facts and laws.

\paragraph{Stage 4: Judgment Generation (Write).}
Finally, the \textit{Document Writing Agent} ($\mathcal{A}_{\mathrm{write}}$) generates the final second-instance judgment. Conditioned on the entire reasoning chain $\mathbf{z} = \{\mathbf{z}_{\mathrm{issue}}, \mathbf{z}_{\mathrm{law}}, \mathbf{z}_{\mathrm{verdict}}\}$, the agent produces the document token-by-token:
\begin{equation}
\begin{aligned}
P(\mathbf{y} \mid \mathbf{x}, \mathbf{z})
&= \prod_{t=1}^{T} P_{\theta}\bigl(
y_t \mid y_{<t}, \mathbf{d}_{\mathrm{1st}}, \mathbf{e}_{\mathrm{new}}, \\
&\qquad\qquad \mathbf{z}_{\mathrm{issue}}, \mathbf{z}_{\mathrm{law}}, \mathbf{z}_{\mathrm{verdict}}
\bigr)
\end{aligned}
\end{equation}

By explicitly conditioning on $\mathbf{z}$, the framework ensures that the generated text strictly adheres to the identified issues, retrieved laws, and predicted verdict, effectively mitigating hallucination.

\begin{table*}[t]
\centering
\resizebox{\textwidth}{!}{%
\begin{tabular}{lcccccccc} 
\toprule
\multirow{3.5}{*}{\textbf{Model}} &
\multicolumn{2}{c}{\textbf{Standard Metrics}} &
\multicolumn{5}{c}{\textbf{LLM-as-a-Judge Scores}} &
\multirow{2}{*}{\textbf{\makecell{Reversal\\Prediction\\Accuracy}}} \\
\cmidrule(lr){2-3} \cmidrule(lr){4-8}
& ROUGE-L & BERTScore & \makecell{Verdict\\Consistency} & \makecell{Fact\\Consistency} & \makecell{Legal\\Application} & \makecell{Logical\\Reasoning} & Average & \\
\midrule

\rowcolor{orange!20} \multicolumn{9}{c}{\textit{Single Non- Reasoning Agent}} \\
Qwen3-235B-A22B-Instruct-2507 & 0.344 & 0.866 & 2.136 & 2.098 & 2.503 & 2.291 & 2.257 & 61.42\% \\
Qwen3-30B-A3B-Instruct-2507 & 0.360 & 0.863 & 2.144 & 2.322 & 2.504 & 2.120 & 2.273 & 60.43\% \\
Qwen3-8B & 0.313 & 0.840 & 1.564 & 2.077 & 2.151 & 1.469 & 1.815 & 57.92\% \\
Gemini-2.5-Flash-Lite & 0.521 & 0.836 & 1.955 & 2.314 & 2.411 & 1.713 & 2.098 & 57.58\% \\
Gemini-2.5-Flash & \textbf{0.536} & 0.856 & 2.396 & 2.451 & 2.558 & 2.323 & 2.432 & 66.53\% \\

\midrule

\rowcolor{blue!20} \multicolumn{9}{c}{\textit{Single Reasoning Agent}} \\
QwQ-32B & 0.277 & 0.851 & 2.366 & \textbf{2.541} & 2.689 & 2.243 & 2.460 & 60.06\% \\
Qwen3-30B-A3B-Thinking-2507 & 0.288 & 0.836 & 2.213 & 2.373 & 2.606 & 2.000 & 2.298 & 64.81\% \\
DeepSeek-R1-0528-Qwen3-8B & 0.316 & 0.837 & 2.137 & 2.229 & 2.293 & 1.697 & 2.089 & 57.87\% \\
gpt-oss-120b & 0.320 & 0.797 & 1.262 & 2.078 & 2.059 & 1.749 & 1.787 & 41.81\% \\
\midrule

\rowcolor{green!20} \multicolumn{9}{c}{\textit{Single Domain Agent}} \\
DISC-LawLLM & 0.080 & 0.660 & 1.643 & 2.075 & 2.237 & 1.683 & 1.910 & 35.22\% \\
Wisdom Interrogatory & 0.333 & 0.721 & 1.443 & 1.983 & 1.731 & 1.240 & 1.599 & 37.34\% \\
\midrule

\rowcolor{purple!20} \multicolumn{9}{c}{\textit{Multi Agents (Our SLMAS)}} \\
Qwen3-235B-A22B-Instruct-2507 & 0.337 & \textbf{0.878} & 2.353 & 2.472 & \textbf{2.705} & \textbf{2.615} & \textbf{2.536} & \textbf{67.24\%}\\
Qwen3-30B-A3B-Instruct-2507 & 0.353 & 0.834 & \textbf{2.460} & 2.406 & 2.430 & 2.274 & 2.393  & 66.49\%  \\
Qwen3-8B & 0.272 & 0.810 & 1.712 & 2.181 & 2.308 & 1.684 & 1.971  & 61.27\%  \\
\bottomrule
\end{tabular}%
}
\caption{Performance comparison on Appellate Legal Judgment Generation task. \textbf{Bold} indicates best performance. }
\label{fig:experiments}
\vspace{-5mm}
\end{table*}

\section{Experiment}

\subsection{Experiment Settings}

\paragraph{Test Dataset Construction.} We curated 1,000 distinct cases from the original dataset through stratified sampling as test set, consisting of 500 cases where the initial judgment was reversed and 500 cases where it was affirmed. This 1:1 ratio ensures that the evaluation metrics accurately reflect the model's reasoning ability.
\paragraph{Baseline.} We leverage LLMs ranging from tens of billions to hundreds of billions of parameters as baselines. The models include:
(1) General Open-source Models: gpt-oss-120b \citep{agarwal2025gpt} and Qwen3 \citep{yang2025qwen3};
(2) Commercial Close-source Models: Gemini 2.5 \citep{comanici2025gemini};
(3) Legal-specific Models: Domain-adapted models including DISC-LawLLM \citep{yue2023disc} and Wisdom Interrogatory \citep{wisdomInterrogatory}.
(4) Inference-Matched Baselines: To ensure a strictly fair comparison and control for computational budgets, we implement Self-Consistency \citep{wang2022self} and Self-Reflection \citep{renze2024self} on the single-agent models. We explicitly align their generation iterations ($N=4$) with the exact number of LLM calls utilized by our SLMAS.

\paragraph{Implementation Details.} All agents in SLMAS are instantiated using Qwen3 \citep{yang2025qwen3}, with role-specific prompts and tool to enable functional specialization. For reproducibility, we fix the decoding temperature at 0.1 across all trials. To ensure fair comparison, identical prompts are employed, and input sequences are truncated to fit within each model's respective context window.

\subsection{Automatic Evaluation}
\label{sec:automatic_evaluation}

We employ a hybrid framework combining LLM-as-a-Judge (reference-free expert scoring) \citep{li2025generation} and standard reference-based metrics to assess the legal validity of generated judgments.

\paragraph{LLM-as-a-Judge.} We utilize DeepSeek-V3.2 \citep{liu2025deepseek} to score generations on a Likert scale (0--5) based on expert-designed rubrics \citep{joshi2015likert}. The detailed rubrics are located in \textbf{Appendix~\ref{app:eval_details}}. The evaluation encompasses four dimensions:
(1) Verdict Consistency: Checks if the ruling direction (Affirm/Reverse) matches the ground truth and whether the judgment results are the same. A mismatch results in a nullified score ($S_{verdict}=0$).
(2) Fact Consistency: Assesses the reconstruction of facts and handling of evidence.
(3) Legal Application: Verifies the accurate citation of operative statutes.
(4) Logical Reasoning: Evaluates syllogistic rigor regarding appeal grounds.

\paragraph{Standard Metrics.} Complementing the judge model, we report ROUGE-L \citep{lin2004rouge} and BERTScore \citep{zhang2019bertscore} to measure lexical overlap and semantic similarity against reference judgments.

\paragraph{Validation of LLM-as-a-Judge.} To mitigate potential verbosity bias, four legal experts evaluated a random 50-case subset. As detailed in \ref{app:eval_details}, the automated scores demonstrate strong and highly significant correlations with human judgments across all dimensions, empirically validating the reliability of our evaluation framework.

\begin{table*}[!h]
\centering
\tiny
\resizebox{\textwidth}{!}{%
\begin{tabular}{lccccccccccc} 
\toprule
\multicolumn{4}{c}{\multirow{1}{*}{\textbf{Agent Configuration}}} &
\multicolumn{2}{c}{\textbf{Standard Metrics}} &
\multicolumn{5}{c}{\textbf{LLM-as-a-Judge Scores}} &
\multirow{2}{*}{\textbf{\makecell{Reversal\\Prediction\\Accuracy}}} \\

\cmidrule(lr){1-4}\cmidrule(lr){5-6} \cmidrule(lr){7-11}

\makecell{Disp.\\Issues} & \makecell{Legal\\Retr.} & \makecell{Rev.\\Pred.} & \makecell{Doc.\\Gen.} & 
ROUGE-L & BERTScore & 
\makecell{Verd.\\Cons.} & \makecell{Fact\\Cons.} & \makecell{Legal\\App.} & \makecell{Logical\\Reas.} & Average & 
\\
\midrule

\rowcolor{oursbg}
\yes & \yes & \yes & \yes & 
0.272 & 0.810 & 
\textbf{1.712} & \textbf{2.181} & \textbf{2.308} & \textbf{1.684} & \textbf{1.971} & 
\textbf{61.27\%} \\ 

\yes & \no & \yes & \yes & 
0.278 & 0.820 & 
1.694 & 2.064 & 2.085 & 1.644 & 1.872 & 
59.30\% \\ 

\no & \yes & \yes & \yes & 
0.269 & 0.803 & 
1.663 & 2.129 & 2.191 & 1.651 & 1.909 & 
60.41\% \\ 

\no & \no & \yes & \yes & 
\textbf{0.331} & 0.823 & 
1.621 & 2.118 & 2.131 & 1.614 & 1.871 & 
58.05\% \\ 

\no & \no & \no & \yes & 
0.313 & \textbf{0.840} & 
1.564 & 2.077 & 2.151 & 1.469 & 1.815 & 
57.92\% \\ 

\bottomrule
\end{tabular}
}
\caption{We perform the Ablation study using Qwen3-8B. We report Generation Similarity (ROUGE-L and BERTScore), LLM-as-a-Judge Scores (including Verdict Consistency, Fact Consistency, Legal Application and Logical Reasoning), and Reversal Prediction Accuracy. The Average column represents the mean of the LLM-as-a-Judge Scores. Best results are \textbf{bold}.}
\label{tab:ablation_study}
\end{table*}

\subsection{Main Result}

\subsubsection{Effectiveness of the SLMAS Framework}

\textbf{Table \ref{fig:experiments}} demonstrates that the SLMAS framework consistently enhances performance across models of varying scales. Using Qwen3-235B-A22B-Instruct-2507 as a primary example, SLMAS improves the average generation quality score from 2.257 to 2.536 and increases Reversal Prediction Accuracy from 61.42\% to 67.24\%. This performance exceeds that of the proprietary Gemini 2.5 Flash (66.53\%), indicating that open-source models can achieve state-of-the-art results when equipped with effective agentic strategies. The improvement is attributed to the structural decomposition of the legal syllogism. By decoupling issue identification from the drafting process, SLMAS effectively models the judicial workflow, thereby mitigating the hallucinations and logical inconsistencies.

Furthermore, to ensure a strictly fair comparison, we evaluated strong single-agent baselines equipped with Self-Consistency \citep{wang2022self} and Self-Reflection \citep{renze2024self}. We aligned their generation iterations ($N=4$) with the exact number of agent calls utilized by SLMAS. As detailed in Appendix~\ref{app:budget_match}, SLMAS significantly outperforms both inference-matched baselines. 

\subsubsection{Reasoning Capabilities versus Domain Knowledge}
A significant divergence in performance is observed between general-purpose LLMs and domain-specific models. While general models achieve accuracies exceeding 60\%, traditional legal models such as DISC-LawLLM and Wisdom Interrogatory lag significantly, with accuracies between 35\% and 37\%. This disparity suggests that appellate judgment generation relies less on static knowledge retrieval and more on comparative reasoning capabilities. Domain-specific models, trained on static legal corpora, lack the cognitive flexibility required to process dynamic evidentiary conflicts between first and second-instance trials. In contrast, general LLMs, particularly those enhanced with chain-of-thought processes, demonstrate superior adaptability in these logic-intensive scenarios.

\subsubsection{Persistent Challenges in Generation Quality}
Despite the improvements introduced by SLMAS, generating second-instance judgments remains a complex challenge. Although prediction accuracy has reached approximately 67\%, the generation quality scores reveal a persistent gap, with average scores hovering around 2.5 out of 5. The primary bottleneck is identified in the \textit{Logical Reasoning} metric (2.615), where models occasionally fail to construct closed-loop arguments that rigorously connect new evidence to statutory conclusions. This indicates that while instruction tuning improves adherence to format, future advancements may require paradigms specifically designed to enhance long-chain causal reasoning.

\subsubsection{Impact of Decoding Randomness and Framework Stability}
To rigorously verify that the performance gains of SLMAS are not artifacts of decoding randomness, we conducted multiple independent runs (N=3) for both the single-agent and multi-agent configurations using the Qwen3-30B \citep{yang2025qwen3} model. According to the experiment results, the SLMAS framework consistently outperforms the single-agent baseline across all trials. The full per-run statistics are reported in Appendix~\ref{app:robustness}.


\subsection{Ablation Study}

We also conducted an ablation study (presented in \textbf{Table \ref{tab:ablation_study}}).

\subsubsection{Dependency of Retrieval on Issue Identification}
The full system achieves optimal performance with an average score of 1.971 and a reversal prediction accuracy of 61.27\%. A critical finding is the impact of the \textit{Disputed Issues Identification} agent. Removing this module degrades performance not only in reasoning but also in legal application. Without explicit issue definition, the subsequent retrieval module lacks directional focus, leading to the retrieval of statutes that fail to ground the final judgment. This confirms that SLMAS prevents error propagation by ensuring that each stage is conditioned on a verified intermediate state.

\subsubsection{Divergence between N-gram and Semantic Metrics}
The \textit{No Agents} baseline achieves the highest ROUGE-L (0.331) but a mediocre LLM-as-a-Judge score (1.871), exposing the limitations of surface-level metrics in legal generation. By verbatim repeating the first-instance judgment, the baseline inflates lexical overlap while lacking dialectical value. Conversely, SLMAS logically restructures responses to address appeal grounds; this naturally reduces n-gram overlap but significantly enhances \textit{Verdict Consistency} (1.712 vs. 1.621) and \textit{Logical Reasoning}. This discrepancy underscores the necessity of semantic-aware evaluations over traditional overlap metrics for high-level reasoning tasks.

\section{Conclusion}
We present \texttt{AppellateGen}, a benchmark shifting the focus of legal intelligence from judgment prediction to the dialectical complexity of second-instance judgment generation. Comprising 7,351 annotated case pairs, it evaluates model reasoning over conflicting narratives and judicial errors. Our proposed SLMAS framework, which structurally decomposes the judicial workflow, significantly enhances logical consistency and reversal accuracy. Notably, general LLMs equipped with reasoning capabilities outperform domain-specific models, underscoring the primacy of cognitive flexibility over static domain pre-training for this task. While a gap remains between generated outputs and human expertise, \texttt{AppellateGen} establishes a foundation for future research into causal dependency modeling and robust comparative reasoning architectures.

\bibliography{custom}

@article{luo2017learning,
  title={Learning to predict charges for criminal cases with legal basis},
  author={Luo, Bingfeng and Feng, Yansong and Xu, Jianbo and Zhang, Xiang and Zhao, Dongyan},
  journal={arXiv preprint arXiv:1707.09168},
  year={2017}
}

@inproceedings{hu2018few,
  title={Few-shot charge prediction with discriminative legal attributes},
  author={Hu, Zikun and Li, Xiang and Tu, Cunchao and Liu, Zhiyuan and Sun, Maosong},
  booktitle={Proceedings of the 27th international conference on computational linguistics},
  pages={487--498},
  year={2018}
}

@article{chen2019charge,
  title={Charge-based prison term prediction with deep gating network},
  author={Chen, Huajie and Cai, Deng and Dai, Wei and Dai, Zehui and Ding, Yadong},
  journal={arXiv preprint arXiv:1908.11521},
  year={2019}
}

@article{kang2019creating,
  title={Creating auxiliary representations from charge definitions for criminal charge prediction},
  author={Kang, Liangyi and Liu, Jie and Liu, Lingqiao and Shi, Qinfeng and Ye, Dan},
  journal={arXiv preprint arXiv:1911.05202},
  year={2019}
}

@article{li2019mann,
  title={Mann: A multichannel attentive neural network for legal judgment prediction},
  author={Li, Shang and Zhang, Hongli and Ye, Lin and Guo, Xiaoding and Fang, Binxing},
  journal={IEEE Access},
  volume={7},
  pages={151144--151155},
  year={2019},
  publisher={IEEE}
}

@article{nigam2024rethinking,
  title={Rethinking legal judgement prediction in a realistic scenario in the era of large language models},
  author={Nigam, Shubham Kumar and Deroy, Aniket and Maity, Subhankar and Bhattacharya, Arnab},
  journal={arXiv preprint arXiv:2410.10542},
  year={2024}
}

@article{guha2023legalbench,
  title={Legalbench: A collaboratively built benchmark for measuring legal reasoning in large language models},
  author={Guha, Neel and Nyarko, Julian and Ho, Daniel and R{\'e}, Christopher and Chilton, Adam and Chohlas-Wood, Alex and Peters, Austin and Waldon, Brandon and Rockmore, Daniel and Zambrano, Diego and others},
  journal={Advances in neural information processing systems},
  volume={36},
  pages={44123--44279},
  year={2023}
}

@inproceedings{ji2023towards,
  title={Towards mitigating LLM hallucination via self reflection},
  author={Ji, Ziwei and Yu, Tiezheng and Xu, Yan and Lee, Nayeon and Ishii, Etsuko and Fung, Pascale},
  booktitle={Findings of the Association for Computational Linguistics: EMNLP 2023},
  pages={1827--1843},
  year={2023}
}

@inproceedings{hong2023metagpt,
  title={MetaGPT: Meta programming for a multi-agent collaborative framework},
  author={Hong, Sirui and Zhuge, Mingchen and Chen, Jonathan and Zheng, Xiawu and Cheng, Yuheng and Wang, Jinlin and Zhang, Ceyao and Wang, Zili and Yau, Steven Ka Shing and Lin, Zijuan and others},
  booktitle={The Twelfth International Conference on Learning Representations},
  year={2023}
}

@article{li2023camel,
  title={Camel: Communicative agents for" mind" exploration of large language model society},
  author={Li, Guohao and Hammoud, Hasan and Itani, Hani and Khizbullin, Dmitrii and Ghanem, Bernard},
  journal={Advances in Neural Information Processing Systems},
  volume={36},
  pages={51991--52008},
  year={2023}
}

@article{huang2025appealcase,
  title={AppealCase: A Dataset and Benchmark for Civil Case Appeal Scenarios},
  author={Huang, Yuting and Guo, Meitong and Wu, Yiquan and Li, Ang and Liu, Xiaozhong and Yin, Keting and Sun, Changlong and Wu, Fei and Kuang, Kun},
  journal={arXiv preprint arXiv:2505.16514},
  year={2025}
}

@article{cui2023chatlaw,
  title={Chatlaw: A multi-agent collaborative legal assistant with knowledge graph enhanced mixture-of-experts large language model},
  author={Cui, Jiaxi and Ning, Munan and Li, Zongjian and Chen, Bohua and Yan, Yang and Li, Hao and Ling, Bin and Tian, Yonghong and Yuan, Li},
  journal={arXiv preprint arXiv:2306.16092},
  year={2023}
}

@article{yuan2026multi,
  title={A multi-agent framework with legal event logic graph for multi-defendant legal judgment prediction},
  author={Yuan, Weikang and Song, Kaisong and Jiang, Zhuoren and Cao, Junjie and Zhang, Yujie and Liu, Chengyuan and Lin, Jun and Zhang, Ji and Kuang, Kun and Liu, Xiaozhong},
  journal={Information Processing \& Management},
  volume={63},
  number={1},
  pages={104319},
  year={2026},
  publisher={Elsevier}
}

@article{liu2025deepseek,
  title={Deepseek-v3. 2: Pushing the frontier of open large language models},
  author={Liu, Aixin and Mei, Aoxue and Lin, Bangcai and Xue, Bing and Wang, Bingxuan and Xu, Bingzheng and Wu, Bochao and Zhang, Bowei and Lin, Chaofan and Dong, Chen and others},
  journal={arXiv preprint arXiv:2512.02556},
  year={2025}
}

@article{yang2025qwen3,
  title={Qwen3 technical report},
  author={Yang, An and Li, Anfeng and Yang, Baosong and Zhang, Beichen and Hui, Binyuan and Zheng, Bo and Yu, Bowen and Gao, Chang and Huang, Chengen and Lv, Chenxu and others},
  journal={arXiv preprint arXiv:2505.09388},
  year={2025}
}

@article{comanici2025gemini,
  title={Gemini 2.5: Pushing the frontier with advanced reasoning, multimodality, long context, and next generation agentic capabilities},
  author={Comanici, Gheorghe and Bieber, Eric and Schaekermann, Mike and Pasupat, Ice and Sachdeva, Noveen and Dhillon, Inderjit and Blistein, Marcel and Ram, Ori and Zhang, Dan and Rosen, Evan and others},
  journal={arXiv preprint arXiv:2507.06261},
  year={2025}
}

@article{yue2023disc,
  title={Disc-lawllm: Fine-tuning large language models for intelligent legal services},
  author={Yue, Shengbin and Chen, Wei and Wang, Siyuan and Li, Bingxuan and Shen, Chenchen and Liu, Shujun and Zhou, Yuxuan and Xiao, Yao and Yun, Song and Huang, Xuanjing and others},
  journal={arXiv preprint arXiv:2309.11325},
  year={2023}
}

@misc{wisdomInterrogatory, author = {Wu, Yiquan and Liu, Yuhang and Liu, Yifei and Li, Ang and Zhou, Siying and Kuang, Kun}, 
title = {wisdomInterrogatory}, url = {https://github.com/zhihaiLLM/wisdomInterrogatory}, version = {1.0}, date = {2024-03-18}, publisher = {GitHub}, note = {Available at GitHub}, year={2024} }

@inproceedings{lin2004rouge,
  title={Rouge: A package for automatic evaluation of summaries},
  author={Lin, Chin-Yew},
  booktitle={Text summarization branches out},
  pages={74--81},
  year={2004}
}

@article{zhang2019bertscore,
  title={Bertscore: Evaluating text generation with bert},
  author={Zhang, Tianyi and Kishore, Varsha and Wu, Felix and Weinberger, Kilian Q and Artzi, Yoav},
  journal={arXiv preprint arXiv:1904.09675},
  year={2019}
}

@article{katz2024gpt,
  title={Gpt-4 passes the bar exam},
  author={Katz, Daniel Martin and Bommarito, Michael James and Gao, Shang and Arredondo, Pablo},
  journal={Philosophical Transactions of the Royal Society A},
  volume={382},
  number={2270},
  pages={20230254},
  year={2024},
  publisher={The Royal Society}
}

@book{apple1995primer,
  title={A primer on the civil-law system},
  author={Apple, James G and Deyling, Robert P},
  year={1995},
  publisher={Federal Judicial Center}
}

@inproceedings{feng2022legal,
  title={Legal Judgment Prediction: A Survey of the State of the Art.},
  author={Feng, Yi and Li, Chuanyi and Ng, Vincent},
  booktitle={IJCAI},
  pages={5461--5469},
  year={2022}
}

@inproceedings{chalkidis-etal-2022-lexglue,
    title = "{L}ex{GLUE}: A Benchmark Dataset for Legal Language Understanding in {E}nglish",
    author = "Chalkidis, Ilias  and
      Jana, Abhik  and
      Hartung, Dirk  and
      Bommarito, Michael  and
      Androutsopoulos, Ion  and
      Katz, Daniel  and
      Aletras, Nikolaos",
    editor = "Muresan, Smaranda  and
      Nakov, Preslav  and
      Villavicencio, Aline",
    booktitle = "Proceedings of the 60th Annual Meeting of the Association for Computational Linguistics (Volume 1: Long Papers)",
    month = may,
    year = "2022",
    address = "Dublin, Ireland",
    publisher = "Association for Computational Linguistics",
    url = "https://aclanthology.org/2022.acl-long.297/",
    doi = "10.18653/v1/2022.acl-long.297",
    pages = "4310--4330"
}

@article{niklaus2023lextreme,
  title={Lextreme: A multi-lingual and multi-task benchmark for the legal domain},
  author={Niklaus, Joel and Matoshi, Veton and Rani, Pooja and Galassi, Andrea and St{\"u}rmer, Matthias and Chalkidis, Ilias},
  journal={arXiv preprint arXiv:2301.13126},
  year={2023}
}

@inproceedings{chalkidis-etal-2022-fairlex,
    title = "{F}air{L}ex: A Multilingual Benchmark for Evaluating Fairness in Legal Text Processing",
    author = "Chalkidis, Ilias  and
      Pasini, Tommaso  and
      Zhang, Sheng  and
      Tomada, Letizia  and
      Schwemer, Sebastian  and
      S{\o}gaard, Anders",
    editor = "Muresan, Smaranda  and
      Nakov, Preslav  and
      Villavicencio, Aline",
    booktitle = "Proceedings of the 60th Annual Meeting of the Association for Computational Linguistics (Volume 1: Long Papers)",
    month = may,
    year = "2022",
    address = "Dublin, Ireland",
    publisher = "Association for Computational Linguistics",
    url = "https://aclanthology.org/2022.acl-long.301/",
    doi = "10.18653/v1/2022.acl-long.301",
    pages = "4389--4406"
}

@inproceedings{fei-etal-2024-lawbench,
    title = "{L}aw{B}ench: Benchmarking Legal Knowledge of Large Language Models",
    author = "Fei, Zhiwei  and
      Shen, Xiaoyu  and
      Zhu, Dawei  and
      Zhou, Fengzhe  and
      Han, Zhuo  and
      Huang, Alan  and
      Zhang, Songyang  and
      Chen, Kai  and
      Yin, Zhixin  and
      Shen, Zongwen  and
      Ge, Jidong  and
      Ng, Vincent",
    editor = "Al-Onaizan, Yaser  and
      Bansal, Mohit  and
      Chen, Yun-Nung",
    booktitle = "Proceedings of the 2024 Conference on Empirical Methods in Natural Language Processing",
    month = nov,
    year = "2024",
    address = "Miami, Florida, USA",
    publisher = "Association for Computational Linguistics",
    url = "https://aclanthology.org/2024.emnlp-main.452/",
    doi = "10.18653/v1/2024.emnlp-main.452",
    pages = "7933--7962"
}

@article{pipitone2024legalbench,
  title={Legalbench-rag: A benchmark for retrieval-augmented generation in the legal domain},
  author={Pipitone, Nicholas and Alami, Ghita Houir},
  journal={arXiv preprint arXiv:2408.10343},
  year={2024}
}

@inproceedings{dai-etal-2025-laiw,
    title = "{LA}i{W}: A {C}hinese Legal Large Language Models Benchmark",
    author = "Dai, Yongfu  and
      Feng, Duanyu  and
      Huang, Jimin  and
      Jia, Haochen  and
      Xie, Qianqian  and
      Zhang, Yifang  and
      Han, Weiguang  and
      Tian, Wei  and
      Wang, Hao",
    editor = "Rambow, Owen  and
      Wanner, Leo  and
      Apidianaki, Marianna  and
      Al-Khalifa, Hend  and
      Eugenio, Barbara Di  and
      Schockaert, Steven",
    booktitle = "Proceedings of the 31st International Conference on Computational Linguistics",
    month = jan,
    year = "2025",
    address = "Abu Dhabi, UAE",
    publisher = "Association for Computational Linguistics",
    url = "https://aclanthology.org/2025.coling-main.716/",
    pages = "10738--10766"
}

@inproceedings{niklaus-etal-2025-lawinstruct,
    title = "{L}aw{I}nstruct: A Resource for Studying Language Model Adaptation to the Legal Domain",
    author = "Niklaus, Joel  and
      Zheng, Lucia  and
      McCarthy, Arya D.  and
      Hahn, Christopher  and
      Rosen, Brian M  and
      Henderson, Peter  and
      Ho, Daniel E.  and
      Honke, Garrett  and
      Liang, Percy  and
      Manning, Christopher D",
    editor = "Chiruzzo, Luis  and
      Ritter, Alan  and
      Wang, Lu",
    booktitle = "Findings of the Association for Computational Linguistics: NAACL 2025",
    month = apr,
    year = "2025",
    address = "Albuquerque, New Mexico",
    publisher = "Association for Computational Linguistics",
    url = "https://aclanthology.org/2025.findings-naacl.7/",
    doi = "10.18653/v1/2025.findings-naacl.7",
    pages = "127--152",
    ISBN = "979-8-89176-195-7"
}

@article{kim2025legalsearchlm,
  title={LegalSearchLM: Rethinking Legal Case Retrieval as Legal Elements Generation},
  author={Kim, Chaeeun and Lee, Jinu and Hwang, Wonseok},
  journal={arXiv preprint arXiv:2505.23832},
  year={2025}
}

@inproceedings{li-etal-2025-legalagentbench,
    title = "{L}egal{A}gent{B}ench: Evaluating {LLM} Agents in Legal Domain",
    author = "Li, Haitao  and
      Chen, Junjie  and
      Yang, Jingli  and
      Ai, Qingyao  and
      Jia, Wei  and
      Liu, Youfeng  and
      Lin, Kai  and
      Wu, Yueyue  and
      Yuan, Guozhi  and
      Hu, Yiran  and
      Wang, Wuyue  and
      Liu, Yiqun  and
      Huang, Minlie",
    editor = "Che, Wanxiang  and
      Nabende, Joyce  and
      Shutova, Ekaterina  and
      Pilehvar, Mohammad Taher",
    booktitle = "Proceedings of the 63rd Annual Meeting of the Association for Computational Linguistics (Volume 1: Long Papers)",
    month = jul,
    year = "2025",
    address = "Vienna, Austria",
    publisher = "Association for Computational Linguistics",
    url = "https://aclanthology.org/2025.acl-long.116/",
    doi = "10.18653/v1/2025.acl-long.116",
    pages = "2322--2344",
    ISBN = "979-8-89176-251-0"
}

@article{chlapanis2025greekbarbench,
  title={GreekBarBench: A Challenging Benchmark for Free-Text Legal Reasoning and Citations},
  author={Chlapanis, Odysseas S and Galanis, Dimitrios and Aletras, Nikolaos and Androutsopoulos, Ion},
  journal={arXiv preprint arXiv:2505.17267},
  year={2025}
}

@article{li2024lexeval,
  title={Lexeval: A comprehensive chinese legal benchmark for evaluating large language models},
  author={Li, Haitao and Chen, You and Ai, Qingyao and Wu, Yueyue and Zhang, Ruizhe and Liu, Yiqun},
  journal={Advances in Neural Information Processing Systems},
  volume={37},
  pages={25061--25094},
  year={2024}
}

@inproceedings{hijazi-etal-2024-arablegaleval,
    title = "{A}rab{L}egal{E}val: A Multitask Benchmark for Assessing {A}rabic Legal Knowledge in Large Language Models",
    author = "Hijazi, Faris  and
      AlHarbi, Somayah  and
      AlHussein, Abdulaziz  and
      Abu Shairah, Harethah  and
      AlZahrani, Reem  and
      AlShamlan, Hebah  and
      Knio, Omar  and
      Turkiyyah, George",
    editor = "Habash, Nizar  and
      Bouamor, Houda  and
      Eskander, Ramy  and
      Tomeh, Nadi  and
      Abu Farha, Ibrahim  and
      Abdelali, Ahmed  and
      Touileb, Samia  and
      Hamed, Injy  and
      Onaizan, Yaser  and
      Alhafni, Bashar  and
      Antoun, Wissam  and
      Khalifa, Salam  and
      Haddad, Hatem  and
      Zitouni, Imed  and
      AlKhamissi, Badr  and
      Almatham, Rawan  and
      Mrini, Khalil",
    booktitle = "Proceedings of the Second Arabic Natural Language Processing Conference",
    month = aug,
    year = "2024",
    address = "Bangkok, Thailand",
    publisher = "Association for Computational Linguistics",
    url = "https://aclanthology.org/2024.arabicnlp-1.20/",
    doi = "10.18653/v1/2024.arabicnlp-1.20",
    pages = "225--249"
}

@article{li2025casegen,
  title={Casegen: A benchmark for multi-stage legal case documents generation},
  author={Li, Haitao and Ye, Jiaying and Hu, Yiran and Chen, Jia and Ai, Qingyao and Wu, Yueyue and Chen, Junjie and Chen, Yifan and Luo, Cheng and Zhou, Quan and others},
  journal={arXiv preprint arXiv:2502.17943},
  year={2025}
}

@book{demarco2013peopleware,
  title={Peopleware: productive projects and teams},
  author={DeMarco, Tom and Lister, Tim},
  year={2013},
  publisher={Addison-Wesley}
}

@article{joshi2015likert,
  title={Likert scale: Explored and explained},
  author={Joshi, Ankur and Kale, Saket and Chandel, Satish and Pal, D Kumar},
  journal={British journal of applied science \& technology},
  volume={7},
  number={4},
  pages={396},
  year={2015},
  publisher={Sciencedomain International}
}

@article{wechsler1977appellate,
  title={The Appellate Jurisdiction of the Supreme Court: Reflections on the Law and the Logistics of Direct Review},
  author={Wechsler, Herbert},
  journal={Wash. \& Lee L. Rev.},
  volume={34},
  pages={1043},
  year={1977},
  publisher={HeinOnline}
}

@inproceedings{zhong2018legal,
  title={Legal judgment prediction via topological learning},
  author={Zhong, Haoxi and Guo, Zhipeng and Tu, Cunchao and Xiao, Chaojun and Liu, Zhiyuan and Sun, Maosong},
  booktitle={Proceedings of the 2018 conference on empirical methods in natural language processing},
  pages={3540--3549},
  year={2018}
}

@inproceedings{liu2023ml,
  title={Ml-ljp: Multi-law aware legal judgment prediction},
  author={Liu, Yifei and Wu, Yiquan and Zhang, Yating and Sun, Changlong and Lu, Weiming and Wu, Fei and Kuang, Kun},
  booktitle={Proceedings of the 46th international ACM SIGIR conference on research and development in information retrieval},
  pages={1023--1034},
  year={2023}
}

@article{xu2020distinguish,
  title={Distinguish confusing law articles for legal judgment prediction},
  author={Xu, Nuo and Wang, Pinghui and Chen, Long and Pan, Li and Wang, Xiaoyan and Zhao, Junzhou},
  journal={arXiv preprint arXiv:2004.02557},
  year={2020}
}

@misc{SPC2021,
  author       = {{Supreme People's Court of the People's Republic of China}},
  title        = {{Judicial Statistics Bulletin of National Courts in 2021}},
  howpublished = {\url{http://gongbao.court.gov.cn/Details/a6c42e26948d3545aea5419fa2beaa.html}},
  year         = {2021},
  note         = {(in Chinese)}
}

@misc{SPC2022,
  author       = {{Supreme People's Court of the People's Republic of China}},
  title        = {{Judicial Statistics Bulletin of National Courts in 2022}},
  howpublished = {\url{http://gongbao.court.gov.cn/Details/20587eaef248beb61ed6596018865c.html}},
  year         = {2022},
  note         = {(in Chinese)}
}

@misc{SPC2023,
  author       = {{Supreme People's Court of the People's Republic of China}},
  title        = {{Judicial Statistics Bulletin of National Courts in 2023}},
  howpublished = {\url{http://gongbao.court.gov.cn/Details/a3e86176b272dc94a05d9cb012c2d5.html}},
  year         = {2023},
  note         = {(in Chinese)}
}

@article{agarwal2025gpt,
  title={gpt-oss-120b \& gpt-oss-20b model card},
  author={Agarwal, Sandhini and Ahmad, Lama and Ai, Jason and Altman, Sam and Applebaum, Andy and Arbus, Edwin and Arora, Rahul K and Bai, Yu and Baker, Bowen and Bao, Haiming and others},
  journal={arXiv preprint arXiv:2508.10925},
  year={2025}
}

@inproceedings{li2025generation,
  title={From generation to judgment: Opportunities and challenges of llm-as-a-judge},
  author={Li, Dawei and Jiang, Bohan and Huang, Liangjie and Beigi, Alimohammad and Zhao, Chengshuai and Tan, Zhen and Bhattacharjee, Amrita and Jiang, Yuxuan and Chen, Canyu and Wu, Tianhao and others},
  booktitle={Proceedings of the 2025 Conference on Empirical Methods in Natural Language Processing},
  pages={2757--2791},
  year={2025}
}

@article{digitale2022tutorial,
  title={Tutorial on directed acyclic graphs},
  author={Digitale, Jean C and Martin, Jeffrey N and Glymour, Medellena Maria},
  journal={Journal of Clinical Epidemiology},
  volume={142},
  pages={264--267},
  year={2022},
  publisher={Elsevier}
}

@misc{LawRefBook2025,
  author       = {RanKKI and {LawRefBook Contributors}},
  title        = {LawRefBook/Laws},
  year         = {2025},
  publisher    = {GitHub},
  journal      = {GitHub repository},
  howpublished = {\url{https://github.com/LawRefBook/Laws}},
  commit       = {2d50496}
}

@article{zhang2025qwen3,
  title={Qwen3 Embedding: Advancing Text Embedding and Reranking Through Foundation Models},
  author={Zhang, Yanzhao and Li, Mingxin and Long, Dingkun and Zhang, Xin and Lin, Huan and Yang, Baosong and Xie, Pengjun and Yang, An and Liu, Dayiheng and Lin, Junyang and others},
  journal={arXiv preprint arXiv:2506.05176},
  year={2025}
}

@article{luneburg1981specially,
  title={Specially Qualified Juries and Expert Nonjury Tribunals: Alternatives for Coping with the Complexities of Modern Civil Litigation},
  author={Luneburg, William V and Nordenberg, Mark A},
  journal={Va. L. Rev.},
  volume={67},
  pages={887},
  year={1981},
  publisher={HeinOnline}
}

@article{bruhl2010deciding,
  title={Deciding When to Decide: How Appellate Procedure Distributes the Costs of Legal Change},
  author={Bruhl, Aaron-Andrew P},
  journal={Cornell L. Rev.},
  volume={96},
  pages={203},
  year={2010},
  publisher={HeinOnline}
}

@article{wang2022self,
  title={Self-consistency improves chain of thought reasoning in language models},
  author={Wang, Xuezhi and Wei, Jason and Schuurmans, Dale and Le, Quoc and Chi, Ed and Narang, Sharan and Chowdhery, Aakanksha and Zhou, Denny},
  journal={arXiv preprint arXiv:2203.11171},
  year={2022}
}

@article{renze2024self,
  title={Self-reflection in llm agents: Effects on problem-solving performance},
  author={Renze, Matthew and Guven, Erhan},
  journal={arXiv preprint arXiv:2405.06682},
  year={2024}
}
\bibliographystyle{colm2026_conference}

\appendix

\section{Limitations and Future Work}
Our study presents a novel benchmark for second-instance legal judgment generation. However, we acknowledge two primary limitations. First, the geographic and linguistic scope of \texttt{AppellateGen} is confined to Mainland China and the Chinese language. This jurisdiction-specific focus, while necessary for legal precision, limits the direct applicability of our models to other legal systems or multilingual contexts. Second, a significant performance bottleneck persists. Although our proposed SLMAS outperforms most of the baselines, the absolute performance remains suboptimal, with a SOTA metric of $67.24\%$ and a LLM-as-a-judge evaluation score of only $2.536$ out of $5$. These results suggest that current paradigms struggle to fully capture the intricate dialectical reasoning required for high-stakes judicial writing. To address these challenges, we look forward to the emergence of foundation models possessing superior legal reasoning capabilities. 

\section{Ethics Statement}
The judgment documents of our dataset are all collected from a publicly accessible platform, China Judgments Online, which is widely used in legal AI research \citep{huang2025appealcase, li2025casegen}. We strictly adhere to the data usage policies and redistribution terms of the source platform, ensuring that our release is confined to de-identified textual content within the boundaries of public access. However, legal data contains sensitive information. To strictly protect the privacy of the individuals involved, we have applied rigorous anonymization preprocessing. All Personally Identifiable Information (PII), such as litigants' names, ID numbers, home addresses, and phone numbers, has been desensitized prior to model training and dataset release. We also acknowledge the potential risks associated with algorithmic bias. As the dataset reflects historical judicial decisions, it may inherently contain biases related to gender, region, or specific crime types. Models trained on this data might perpetuate these biases. Therefore, we emphasize that this dataset and the proposed method are intended for academic research and as assistive tools for legal professionals only. They should never be used as a substitute for human judges or to automate final judicial decisions. We strongly advise future researchers to conduct fairness evaluations when deploying systems based on this dataset.

\section{GenAI Usage Statement}
We clarify the use of generative AI in this study as follows: (1) Research Methodology: As detailed in Sections 3.3 and 5.2, LLMs were utilized for data annotation and as a scoring judge to assess judgment generation. (2) Writing and Coding: LLMs were used as productivity tools to optimize the codebase and refine the linguistic expression of the paper. All final outputs were critically reviewed by the authors.

\section{Background on Appellate (Second-Instance) Review in Civil Law Jurisdictions}
\label{app: appellate}

\begin{figure}[h]
    \centering
    \includegraphics[width=0.8\linewidth]{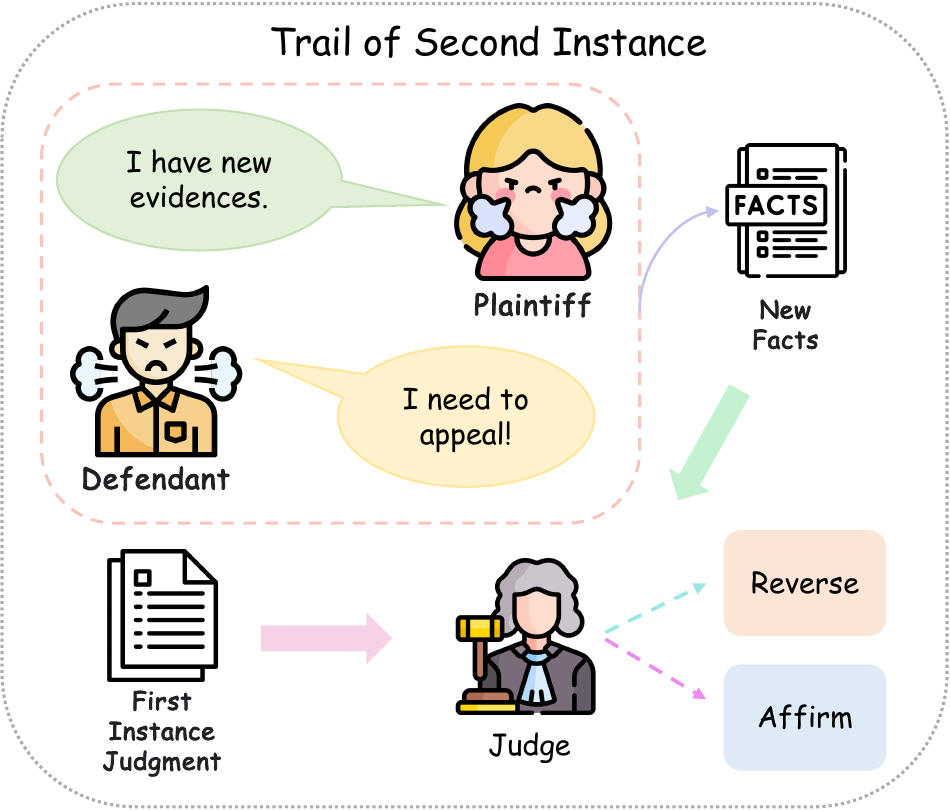}
    \caption{Proceedings of the Second Instance: Following the filing of an appeal by the appellant, the admission of new evidence may result in either the reversal of the verdict or the affirmation of the original judgment.}
    \label{fig:trail}
\end{figure}

The procedural scope of appellate review exhibits fundamental differences between Common Law and Civil Law jurisdictions. In Common Law systems, appellate courts typically exercise deference to the lower court regarding factual determinations, restricting their review primarily to errors of law. Conversely, Civil Law jurisdictions adopt a comprehensive approach where the court of second instance functions as a continuation of the trial process, retaining the authority to re-examine both legal applications and factual findings. This study focuses specifically on the Civil Law framework, utilizing a dataset derived from Chinese legal proceedings.

Within this context, the second-instance trial serves as a critical remedial mechanism. As illustrated in \textbf{Figure \ref{fig:trail}}, the procedure is initiated when a litigant, who may be either the plaintiff or the defendant, files an appeal expressing dissatisfaction with the first-instance judgment. A defining characteristic of this process in China is the admissibility of new evidence. Unlike systems that strictly limit the appellate record, this framework permits the introduction of supplementary facts that were previously unavailable or overlooked.

Consequently, the input for the appellate judge comprises two distinct streams of information: the text of the original first-instance judgment and the set of newly presented evidentiary facts. The presiding judge is tasked with synthesizing these inputs to evaluate the validity of the appeal. This review process necessitates a comparative analysis where the judge assesses whether the new evidence sufficiently contradicts the original findings or if the initial application of the law remains robust. The proceeding culminates in a binary judicial determination. The court may decide to affirm the original judgment if the appeal lacks sufficient merit, effectively affirming the prior ruling. Alternatively, if the combination of the original context and new facts reveals substantive errors, the court issues a reversal, thereby altering the legal verdict. \textbf{Figure \ref{fig:trail}} provides a schematic representation of this workflow, delineating the causal path from the filing of the appeal and the admission of evidence to the final adjudication.

\section{AppellateGen Construction Details}
To construct AppellateGen, we perform two main steps: (1) Data Collection and Legal Instance Matching, and (2) Fine-grained Data Annotation. The construction details are as follows.
\subsection*{Legal Instances Matching}
\label{app: Legal Instances Matching}



\begin{figure}[t]
    \centering
    \includegraphics[width=\columnwidth]{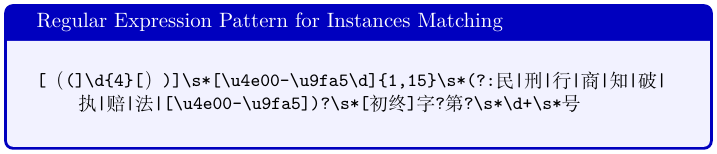} 
    \caption{RE pattern for legal instances matching. This pattern can extract the case identifier of first instance from second-instance judgment document.}
    \label{fig:case number}
\end{figure}

First, we employed high-precision Regular Expressions (RE) to extract unique case identifier of first instance from second-instance judgment document. The specific extraction pattern is detailed in \textbf{Figure \ref{fig:case number}}. Subsequently, we retrieved the corresponding first-instance judgment documents based on these identifiers to construct candidate case pairs.

To ensure the semantic integrity of these pairs, filtering out potential mismatches caused by clerical errors in case numbers, we utilized a LLM as a semantic validator. Specifically, we prompted the LLM to compare the litigants, the cause of action, and the court hierarchy described in both documents, outputting a binary verification score indicating whether they refer to the identical legal dispute. The used prompts are detailed below. 
%
%
\begin{boxL}
You are an experienced expert in reviewing Chinese legal documents, specializing in the filing and correlation of civil, criminal, and administrative case files. You possess exceptional logical analysis and attention to detail.

Your task is to determine whether two given legal documents (a first-instance judgment and a second-instance judgment) belong to the same case's proceedings (i.e., whether the second instance appealed against the first-instance judgment).

\textbf{Please follow these steps for logical reasoning, without directly providing a conclusion:}

\textbf{1. Extract Key Information:}

From the first-instance judgment: Extract the first-instance case number, plaintiff/defendant's name (or title), and cause of action.

From the second-instance judgment: Extract the second-instance case number, appellant/respondent's name, and the original trial case number cited in the text.

\textbf{2. Compare Relevance:}
\begin{itemize}
    \item Case Number Verification: Check the "Case Origin" or "Original Trial Circumstances" section...
    \item Verification of Parties: Does the appellant/appellee in the second instance correspond to the plaintiff/defendant...
    \item Verification of Facts: Are the core facts of the dispute consistent?
\end{itemize}

\textbf{3. Elimination of Interference:}
Pay attention to distinguishing between "related cases"...

\textbf{Output Format:}
Please strictly follow the following JSON format for outputting results:

\{\\
    "is\_same\_case": true/false,\\
    "confidence\_score": "0-10",\\
    "reasoning": ""\\
\}
\end{boxL}

\subsection*{Automated Annotation Pipeline}
\label{app: Automated Annotation Pipeline}
To guarantee the reliability of annotations within complex legal contexts, we established a robust three-stage pipeline:

(1) Structural Decomposition: We employed customized regular expressions to partition unstructured judgment documents into distinct functional units, specifically the \textit{Factual Description} and \textit{Judicial Reasoning} sections. This segmentation mitigates input noise and directs the model's attention toward contextually relevant information.

(2) LLM-Driven Annotation: We used DeepSeek-V3.2 \citep{liu2025deepseek} to generate preliminary annotations. To bolster reasoning transparency and label accuracy, we incorporated a Chain-of-Thought (CoT) prompting strategy (Detailed in Appendix \ref{app: annotation}). This approach compels the model to articulate the underlying legal rationale and calculate a self-assessed confidence score ($\gamma$) prior to determining the final label.

(3) Human-in-the-Loop Verification: To ensure a rigorous ground truth, we instituted a confidence-aware review protocol. Instances with a model confidence below a strict threshold ($\gamma < 0.8$)—comprising approximately 5\% of the dataset—were flagged for manual adjudication by legal experts. For this subset, the Inter-Annotator Agreement (IAA) achieved a Cohen's Kappa ($\kappa$) of 0.827. This "near perfect" agreement between human experts and our pipeline confirms the high reliability of our automated annotations.

\subsection*{Fine-grained Data Annotation}
\label{app: annotation}
Here is the prompt template for LLM-driven annotation.

\begin{boxL}
Please act as a professional expert in Chinese legal document analysis. I will provide the full text of the [First Instance Judgment] and the [Second Instance Judgment] for the same case.

\textbf{Your tasks are:}

\begin{enumerate}
    \item \textbf{Deep Analysis and Extraction:} Compare the two documents and extract key information.
    \item \textbf{Formatted Output:} Integrate all analysis results into a strict JSON object.
\end{enumerate}

Please output strictly according to the following JSON structure and do not include Markdown tags:

\begin{ttfamily}
\{ \\
\hspace*{-1em} "reasoning\_trace": "Your step-by-step analysis here...", \\
\hspace*{1em} "confidence\_score": float, \\
\hspace*{1em} "is\_reversal": bool, \\
\hspace*{1em} "first\_instance": \{ \\
\hspace*{2em} "facts": "string", \\
\hspace*{2em} "disputed\_issues": "string", \\
\hspace*{2em} "legal\_articles": ["string"], \\
\hspace*{2em} "judgment": "string" \\
\hspace*{1em} \}, \\
\hspace*{1em} "second\_instance": \{ \\
\hspace*{2em} "new\_facts": "string", \\
\hspace*{2em} "disputed\_issues": "string", \\
\hspace*{2em} "legal\_articles": ["string"], \\
\hspace*{2em} "judgment": "string" \\
\hspace*{1em} \}, \\
\hspace*{1em} "reason\_for\_reversal": "string" \\
\}
\end{ttfamily}\\
Here are the explanations of the following JSON schemas.\\
\begin{itemize}
    \item \textbf{is\_reversal}: A boolean value indicating whether the
    second-instance judgment reverses.

    \item \textbf{first\_instance}: A structured summary of the first-instance
    judgment, including factual findings, disputed issues, cited legal articles,
    and the final ruling.
    \begin{itemize}
        \item \textbf{facts}: Core facts established in the first instance.
        \item \textbf{disputed\_issues}: Key disputes addressed by the court.
        \item \textbf{legal\_articles}: Statutory provisions relied upon.
        \item \textbf{judgment}: Outcome of the first-instance ruling.
    \end{itemize}

    \item \textbf{second\_instance}: A structured summary of the appellate
    judgment, emphasizing changes relative to the first instance.
    \begin{itemize}
        \item \textbf{new\_facts}: Newly identified facts or evidence.
        \item \textbf{disputed\_issues}: Disputed issues considered on appeal.
        \item \textbf{legal\_articles}: Controlling statutory provisions.
        \item \textbf{judgment}: Outcome of the second-instance ruling.
    \end{itemize}

    \item \textbf{reason\_for\_reversal}: Legal grounds for reversal; left empty if
    the judgment is affirmed.
\end{itemize}

Please begin analyzing the document content I provide.
\end{boxL}

\section{SLMAS details}
\label{app: SLMAS}

\begin{wraptable}{r}{0.5\textwidth}
    \centering
    \vspace{-1.5em} 
    \label{tab:retrieval_performance}
    \begin{tabular}{lccc}
        \toprule
        \textbf{Model} & \textbf{MRR@100} & \textbf{P@10} & \textbf{R@10} \\
        \midrule
        TF-IDF & 0.322 & 0.103 & 0.317 \\
        BM25   & 0.445 & 0.090 & 0.277 \\
        \rowcolor{oursbg} 
        Ours & \textbf{0.752} & \textbf{0.144} & \textbf{0.762} \\
        \bottomrule
    \end{tabular}
    \caption{Performance comparison of retrieval methods.}
\end{wraptable}

We construct a specialized legal knowledge base comprising statutes enacted post-2021, leveraging the open-source \texttt{LawRefBook} toolkit \citep{LawRefBook2025}. A primary advantage of this repository is its seamless synchronization with the authoritative \textit{National Laws and Regulations Database of China}\footnote{This platform serves as the official legislative database maintained by the Ministry of Justice of the People’s Republic of China (\url{https://flk.npc.gov.cn}).}. To facilitate efficient semantic matching, we encode the textual content of each legal article using Qwen3-Embedding \citep{zhang2025qwen3}.

During the retrieval phase, the system first executes a coarse-grained search to extract the top-10 most relevant candidates. To maximize precision, an LLM subsequently acts as a fine-grained selector, filtering these candidates to pinpoint the specific legal articles and judicial interpretations pertinent to the dispute.

As detailed in Table \ref{tab:retrieval_performance}, our Legal Retrieval Agent achieves a Recall@10 of 0.762, significantly outperforming traditional lexical baselines such as BM25 (0.277) and TF-IDF (0.317). Achieving this high recall is essential for providing accurate context to the downstream reasoning agents.

\section{Evaluation Details}
\label{app:eval_details}

In this section, we provide the specific scoring rubrics and the prompt template used for the LLM-as-a-Judge evaluation described in \textbf{Section~\ref{sec:automatic_evaluation}}.

\subsubsection*{Scoring Rubrics}
We designed fine-grained criteria to guide the evaluator (DeepSeek-V3.2) in scoring the generated judgments. The scoring scale ranges from 0 to 5 across four specific dimensions. The detailed rubrics are presented in \textbf{Table~\ref{tab:full_rubrics}}.

\subsubsection*{Evaluation Prompt Template}
We employed a zero-shot prompting strategy with detailed instructions to ensure the LLM evaluator acts as an impartial judge. The full prompt used in our experiments is presented below.

\begin{boxL}

\textbf{System Role:} 
You are a senior legal expert in the Chinese judicial system, specializing in appellate review. Your task is to evaluate a second-instance judgment.

\textbf{Evaluation Task:}
Please score the generated output from 0 to 5 on the following dimensions based on the provided rubrics.

\textbf{1. Verdict Consistency (0-5):}
Does the model make the correct decision (Affirm vs. Reverse)? 
\textit{CRITICAL RULE:} If the ruling direction contradicts the Ground Truth (e.g., model affirms while truth reverses), you \textbf{MUST} give 0 points.

\textbf{2. Fact Consistency (0-5):}
Does the model accurately describe the facts? Crucially, does it capture the "New Evidence" introduced in the second instance? If the model ignores new evidence that leads to a reversal, score low.

\textbf{3. Legal Application (0-5):}
Are the cited statutes correct and relevant? Does it cite the operative substantive law rather than just procedural descriptions?

\textbf{4. Logical Reasoning (0-5):}
Is the reasoning coherent? Does it correctly refute or support the appeal grounds? For reversal cases, does it clearly explain \textit{why} the first instance was wrong?

\textbf{Output Format:}
Please output a strictly valid JSON object:\\
\{\\
  "verdict\_score": int,\\    
  "fact\_score": int,\\       
  "law\_score": int,\\        
  "logic\_score": int,\\      
  "justification": "A brief explanation of the scoring..."\\
\}
\end{boxL}

\subsubsection*{Validation of LLM-as-a-Judge}

\begin{wraptable}{r}{0.48\textwidth}
    \vspace{-15pt} 
    \centering
    \footnotesize 
    \begin{tabular}{llcc}
    \toprule
    \textbf{Dimension} & \textbf{Metric} & \textbf{Score} & \textbf{$p$-value} \\
    \midrule
    Rev. Pred. & Cohen's $\kappa$ & 0.452 & - \\
    \midrule
    Verdict Cons. & Spearman's $\rho$ & 0.770 & $<0.001$ \\
    Fact Cons. & Spearman's $\rho$ & 0.453 & $<0.001$ \\
    Legal App. & Spearman's $\rho$ & 0.630 & $<0.001$ \\
    Logical Reas. & Spearman's $\rho$ & 0.710 & $<0.001$ \\
    \bottomrule
    \end{tabular}
    \caption{Alignment between LLM-as-a-Judge and human experts.}
    \label{tab:human_validation}
    \vspace{-10pt} 
\end{wraptable}

To address potential verbosity bias and ensure evaluative integrity, four legal experts conducted an independent review of a random 50-case subset. As detailed in Table \ref{tab:human_validation}, the alignment between the LLM-as-a-Judge and human experts was robust, with coefficients ranging from 0.4 to 0.8. Specifically, the model demonstrated substantial agreement in categorical prediction (Cohen’s $\kappa = 0.452$) and strong-to-very-strong correlations across all qualitative dimensions ($p < 0.001$), confirming that the LLM's judgment aligns closely with expert legal reasoning. This high degree of correlation underscores the reliability of using automated prompts for large-scale judicial analysis.

\section{Additional Experimental Results}
\label{app:additional_results}

This appendix provides supplementary evidence for two claims made in the main text: (1) the superiority of SLMAS is not attributable to a larger inference budget, and (2) its performance gains are stable across independent runs.

\subsection*{Comparison Under Matched Inference Budgets}
\label{app:budget_match}
To ensure that the gains of SLMAS are not merely caused by a larger number of inference calls, we report a controlled comparison under matched inference budgets in \textbf{Table} \ref{tab:inference_budget}. In particular, the iterative single-agent baselines use the same number of LLM calls as SLMAS, so that this comparison isolates the contribution of the SOP-based multi-agent workflow itself. Notably, while single-pass and iterative baselines occasionally achieve superficially higher N-gram scores (ROUGE/BERTScore) primarily through verbatim copying of input facts, they struggle severely on semantic reasoning. For instance, SLMAS achieves a Legal Application score of 2.430, compared to just 2.067 for Self-Consistency. This divergence decisively proves that the gains of SLMAS are intrinsically tied to the SOP workflow's ability to drive substantive legal reasoning.

\subsection*{Robustness Across Independent Runs}
\label{app:robustness}
To complement the summary in the main text, we further provide the full results of three independent runs for both the single-agent baseline and SLMAS. As shown in \textbf{Table} \ref{tab:robustness_runs}, the SLMAS framework consistently outperforms the single-agent baseline across all trials. Specifically, the multi-agent setup achieves a stable Reversal Prediction Accuracy of 63.26\% ± 3.50\% and an Average Judge Score of 2.2608 ± 0.0943, compared to 57.79\% ± 3.13\% and 2.1865 ± 0.0746 for the single agent.

\newpage
\begin{table}[t]
\centering
\resizebox{\textwidth}{!}{%
\begin{tabular}{lcccccccc} 
\toprule
\multirow{3.5}{*}{\textbf{Method}} &
\multicolumn{2}{c}{\textbf{Standard Metrics}} &
\multicolumn{5}{c}{\textbf{LLM-as-a-Judge Scores}} &
\multirow{2}{*}{\textbf{\makecell{Reversal\\Prediction\\Accuracy}}} \\
\cmidrule(lr){2-3} \cmidrule(lr){4-8}
& ROUGE-L & BERTScore & \makecell{Verdict\\Consistency} & \makecell{Fact\\Consistency} & \makecell{Legal\\Application} & \makecell{Logical\\Reasoning} & Average & \\
\midrule

\rowcolor{orange!20} \multicolumn{9}{c}{\textit{Single-Pass Baseline ($N=1$)}} \\
Standard Prompting & \textbf{0.357} & \textbf{0.861} & 2.088 & 2.197 & 2.203 & 2.098 & 2.147 & 59.54\% \\
\midrule

\rowcolor{blue!20} \multicolumn{9}{c}{\textit{Inference-Matched Baselines ($N=4$)}} \\
Self-Reflection & 0.322 & 0.848 & 1.750 & 1.911 & 1.899 & 1.751 & 1.828 & 53.95\% \\
Self-Consistency & 0.328 & 0.846 & 1.478 & 1.981 & 2.067 & 1.829 & 1.839 & 45.79\% \\
\midrule

\rowcolor{purple!20} \multicolumn{9}{c}{\textit{Multi-Agent Framework (Our SLMAS, $N=4$)}} \\
SLMAS (Ours) & 0.353 & 0.834 & \textbf{2.460} & \textbf{2.406} & \textbf{2.430} & \textbf{2.274} & \textbf{2.393} & \textbf{66.49\%} \\
\bottomrule
\end{tabular}%
}
\caption{Performance comparison under controlled inference budgets using the Qwen3-30B model. \textbf{Bold} indicates best performance in reasoning validity. Note that baseline models achieve superficially higher N-gram scores (ROUGE/BERTScore) primarily by verbatim copying of input facts, whereas SLMAS excels in semantic legal reasoning despite lower lexical overlap.}
\label{tab:inference_budget}
\end{table}

\begin{table}[t]
\centering
\resizebox{\textwidth}{!}{%
\begin{tabular}{ll cccccccc}
\toprule
\multirow{2}{*}{\textbf{Configuration}} & \multirow{2}{*}{\textbf{Run}} & \multicolumn{2}{c}{\textbf{Standard Metrics}} & \multicolumn{5}{c}{\textbf{LLM-as-a-Judge Scores}} & \multirow{2}{*}{\textbf{\makecell{Rev.\\Acc.}}} \\
\cmidrule(lr){3-4} \cmidrule(lr){5-9}
& & BERTScore & ROUGE-L & \makecell{Verd.\\Cons.} & \makecell{Fact\\Cons.} & \makecell{Legal\\App.} & \makecell{Logical\\Reas.} & Average & \\
\midrule

\multirow{4}{*}{\makecell[l]{Single-Agent\\(Baseline)}}
& Run 1 & 0.8612 & 0.3572 & 2.0884 & 2.1968 & 2.2026 & 2.0984 & 2.1466 & 59.54\% \\
& Run 2 & 0.8620 & 0.3520 & 1.8590 & 2.1398 & 2.2996 & 2.0639 & 2.0906 & 53.40\% \\
& Run 3 & 0.8630 & 0.3600 & 2.1440 & 2.3220 & 2.2040 & 2.1200 & 2.1975 & 60.43\% \\
\rowcolor{summarybg} 
& Mean$\pm$Std & 0.8621$\pm$0.0007 & 0.3564$\pm$0.0033 & 2.0305$\pm$0.1234 & 2.2195$\pm$0.0761 & 2.2354$\pm$0.0834 & 2.0941$\pm$0.0231 & 2.1865$\pm$0.0746 & 57.79\%$\pm$3.13\% \\
\midrule

\multirow{4}{*}{\makecell[l]{Multi-Agent\\(Our SLMAS)}}
& Run 1 & 0.8278 & 0.3133 & 2.3260 & 2.2076 & 2.3956 & 2.1234 & 2.2632 & 64.89\% \\
& Run 2 & 0.8288 & 0.3018 & 2.3300 & 2.1733 & 2.4705 & 2.0327 & 2.2516 & 58.39\% \\
& Run 3 & 0.8340 & 0.3530 & 2.4600 & 2.4060 & 2.4300 & 2.2740 & 2.3925 & 66.49\% \\
\rowcolor{oursbg} 
& Mean$\pm$Std & 0.8302$\pm$0.0027 & 0.3227$\pm$0.0219 & \textbf{2.3720$\pm$0.0622} & \textbf{2.2623$\pm$0.1026} & \textbf{2.4300$\pm$0.1169} & \textbf{2.1434$\pm$0.0995} & \textbf{2.2608$\pm$0.0943} & \textbf{63.26\%$\pm$3.50\%} \\
\bottomrule
\end{tabular}%
}
\vspace{2mm} 
\caption{Robustness analysis of the Qwen3-30B model over three independent runs. We report the exact scores for each run along with the overall mean and standard deviation. The \colorbox{oursbg}{blue highlighted row} demonstrates that our SLMAS framework achieves superior and stable legal reasoning performance (indicated by higher judge scores and reversal accuracy) while naturally reducing surface-level text copying (indicated by lower N-gram overlaps) compared to the single-agent baseline.}
\label{tab:robustness_runs}
\end{table}

\begin{table*}[t]
    \centering
    \small
    \renewcommand{\arraystretch}{1.4}
    \begin{tabularx}{\textwidth}{l|c|X}
    \toprule
    \textbf{Dimension} & \textbf{Score} & \textbf{Criteria Description} \\
    \midrule
    
    \multirow{5}{*}{\makecell[l]{\textbf{1. Verdict}\\\textbf{Consistency}}} 
    & \cellcolor{red!10}\textbf{0} & \textbf{Incorrect Ruling Direction.} The generated judgment contradicts the ground truth (e.g., affirming instead of reversing). \textit{*Strict Penalty*} \\
    \cline{2-3}
    & 1--2 & \textbf{Ambiguous/Incomplete.} The direction is unclear, contradictory, or missing the operative part entirely. \\
    \cline{2-3}
    & 3--4 & \textbf{Direction Correct but Flawed.} The general ruling (Affirm/Reverse) is correct, but specific execution orders (e.g., amounts, specific behaviors) are imprecise or missing. \\
    \cline{2-3}
    & 5 & \textbf{Perfect Match.} The operative part is identical in legal effect and scope to the ground truth, fully resolving the dispute. \\
    \hline
    
    \multirow{5}{*}{\makecell[l]{\textbf{2. Fact}\\\textbf{Consistency}}} 
    & 0--1 & \textbf{Hallucination.} Fabricates facts not present in the input or contradicts the core timeline of the case. \\
    \cline{2-3}
    & 2--3 & \textbf{Missing New Evidence.} Accurately recaps the first-instance facts but fails to identify or incorporate the "New Facts" introduced during the appeal. \\
    \cline{2-3}
    & 4 & \textbf{Accurate.} Captures both original facts and new evidence, with minor omissions in non-critical details. \\
    \cline{2-3}
    & 5 & \textbf{Comprehensive.} Perfectly reconstructs the factual basis, highlighting the contradiction between new evidence and original findings. \\
    \hline

    \multirow{4}{*}{\makecell[l]{\textbf{3. Legal}\\\textbf{Application}}} 
    & 0--1 & \textbf{Erroneous Citation.} Cites non-existent statutes or applies laws completely irrelevant to the case cause. \\
    \cline{2-3}
    & 2--3 & \textbf{Incomplete Citation.} Misses the key operative article that dictates the reversal/affirmation, citing only procedural rules. \\
    \cline{2-3}
    & 4--5 & \textbf{Precise Citation.} Correctly identifies and applies the specific articles (e.g., Civil Code Art. X) used in the ground truth. \\
    \hline

    \multirow{5}{*}{\makecell[l]{\textbf{4. Logical}\\\textbf{Reasoning}}} 
    & 0--1 & \textbf{Incoherent.} The reasoning is logically fractured or fails to address the appellant's grievances entirely. \\
    \cline{2-3}
    & 2--3 & \textbf{Mechanical Mapping.} Performs simple fact-to-law mapping but lacks comparative analysis. Misses the causal link between new evidence and the verdict change. \\
    \cline{2-3}
    & 4 & \textbf{Strong Syllogism.} Explicitly discusses why the first-instance judgment was erroneous based on the points of contention. \\
    \cline{2-3}
    & 5 & \textbf{Expert Dialectics.} Demonstrates a clear "review-and-correct" logic, forming a watertight argument that mirrors a human judge's opinion. \\
    \bottomrule
    \end{tabularx}
    \caption{Detailed Scoring Rubrics for Automatic Evaluation. The evaluator scores each generated judgment from 0 to 5 based on these criteria. Note that a score of 0 in Verdict Consistency is strictly enforced if the ruling direction is wrong.}
    \label{tab:full_rubrics}
\end{table*}





\end{document}